\documentclass[conference]{IEEEtran}
\IEEEoverridecommandlockouts
\usepackage[UglyObsolete]{diagrams}
\usepackage{amsmath}
\usepackage{amssymb}
\usepackage{graphics}
\usepackage{latexsym}
\newtheorem{Main}{Main Lemma}
\newtheorem{Proposition}{Proposition}

\newtheorem{Definition}{Definition}
\newtheorem{Example}{Example}
\newtheorem{Algorithm}{Algorithm}
\def\Carrow{\overset{\mathcal{C}}{\longrightarrow}}

\begin{document}

\title{
Fast Erasure-and-Error Decoding and Systematic Encoding
of a Class of Affine Variety Codes{\huge$\,{}^\ast$}
\thanks{${}^\ast\,$Presented at {\it The 34th Symposium on Information Theory and Its Applications (SITA2011),} pp.405--410, Ousyuku, Iwate, Japan, Nov.~29--Dec.~2, 2011.}}

\author{
\IEEEauthorblockN{Hajime Matsui}
\IEEEauthorblockA{Toyota Technological Institute\\
Hisakata, Tenpaku, Nagoya 468--8511, Japan\\
Email: matsui@toyota-ti.ac.jp}
}

\maketitle

\begin{abstract}
In this paper, a lemma in algebraic coding theory is established, which is frequently appeared in the encoding and decoding for algebraic codes such as Reed--Solomon codes and algebraic geometry codes.
This lemma states that two vector spaces, one corresponds to information symbols and the other is indexed by the support of Gr\"obner basis, are canonically isomorphic, and moreover, the isomorphism is given by the extension through linear feedback shift registers from Gr\"obner basis and discrete Fourier transforms.
Next, the lemma is applied to fast unified system of encoding and decoding erasures and errors in a certain class of affine variety codes.

\end{abstract}

\IEEEpeerreviewmaketitle

\section{Introduction}
Despite many researches have been done for both of encoding and erasure-and-error decoding, any relation between them have never been found so far except maximum distance separable (MDS) codes, which satisfy $d-1=n-k$, where $n$, $k$, and $d$ is the code length, dimension, and the minimum distance of the code, respectively.
Since the correctable numbers $u$ of erasures and $t$ of errors satisfy $2t+u\le d-1$, it is obvious for MDS codes that, if $t=0$, then erasure-only decoding can determine $n-k$ redundant symbols, that is, systematic encoding is done by erasure-only decoding.
In general, algebraic geometry codes are not MDS codes; there has never been known no such case where erasure decoding can undertake systematic encoding.

In this paper, we first establish a lemma that is essential in algebraic coding theory.
We observe that almost all manipulations on algebraic Goppa codes such as encoding and decoding are described in terms of the lemma, which provides an isomorphism between a certain pair of vector spaces over a finite field.
The isomorphism of the lemma is written by the combined map of the extension by the linear recurrence relation from a Gr\"obner basis and $N$-dimensional inverse discrete Fourier transform (IDFT).
Next, the lemma is applied to a class of affine variety codes \cite{Fitzgerald-Lax},\cite{Miura affine}, which are essentially same as order domain codes or evaluation codes \cite{Andersen-Geil},\cite{generic},\cite{BMS}, and enables us to decode efficiently erasures and errors.
Finally, we notice that, in a class of affine variety codes, systematic encoding can be viewed as a certain type of erasure-only decoding.

The rest of this paper is organized as follows.
In Section \ref{Notations}, we prepare notations.
In Section \ref{Main lemma}, we state the lemma.
In Subsection \ref{Fourier-type}, we generalize discrete Fourier transforms (DFTs) from on $\left(\mathbb{F}_q^{\times}\right)^N$ into on $\mathbb{F}_q^N$.
In Subsection \ref{vector spaces}, two vector spaces are defined via Gr\"obner basis.
In Subsection \ref{Isomorphic map}, we give an isomorphism between the vector spaces.
In Section \ref{Application}, we apply the lemma.
In Subsection \ref{arbitrary subset}, we construct affine variety codes in terms of the lemma.
In Subsection \ref{DFT encoding}, an erasure-and-error decoding algorithm and its relation with systematic encoding is described.
In Section \ref{Estimation}, the number of finite-field operations in our algorithm is estimated.
Section \ref{Conclusion} concludes the paper.

\section{Notations
\label{Notations}}

Throughout this paper, $\mathbb{N}_0$ is the set of non-negative integers and $\alpha$ is a fixed primitive element of finite field $\mathbb{F}_q$, where $q$ is a prime power.
For $a,b\in\mathbb{N}_0$ with $a\le b$, let $[a,b]:=\left\{a,a+1,\cdots,b\right\}$.
For two sets $A$ and $B$, $A\backslash B$ is defined as $\{u\in A\,|\,u\not\in B\}$.
For arbitrary finite set $S$, let $V_S:=\left.\left\{\left(v_s\right)_S\,\right|\,s\in S,\,v_s\in\mathbb{F}_q\right\}$ denote a vector space over $\mathbb{F}_q$ whose componensts are indexed by $S$.
For any arbitrary subset $R\subset S$, the vector space $V_R$ is considered as a subspace of $V_S$ by $V_R=\left.\left\{\left(v_s\right)_S\in V_S\,\right|\,v_s=0\mbox{ for all }s\in S\backslash R\right\}$.

\section{Main lemma
\label{Main lemma}}

\subsection{Fourier-type transforms on $\mathbb{F}_q^N$
\label{Fourier-type}}

Let $N$ be a positive integer and let
\begin{align*}
&A\;:=[0,q-1]^N\\
&=\left\{\left.\underline{a}=\left(a_1,\cdots,a_N\right)\,\right|\,a_1,\cdots,a_N\in[0,q-1]\,\right\},\\
&\Omega\;:=\mathbb{F}_q^N=\left\{\underline{\omega}=\left(\omega_1,\cdots,\omega_N\right)\,\left|\,\omega_1,\cdots,\omega_N\in\mathbb{F}_q\right\}\right..
\end{align*}
In this subsection, Fourier-type transforms are defined as maps between two vector spaces, which are isomorphic to $\mathbb{F}_q^N$,
\begin{align*}
V_A&:=\left\{\left(h_{\underline{a}}\right)_A\,\left|\,\underline{a}\in A,\,h_{\underline{a}}\in\mathbb{F}_q\right.\right\},\\
V_{\Omega}&:=\left\{\left.\left(c_{\underline{\omega}}\right)_{\Omega}\,\right|\,\underline{\omega}\in\Omega,\,c_{\underline{\omega}}\in\mathbb{F}_q\right\}.
\end{align*}
For $\left(c_{\underline{\omega}}\right)_\Omega\in V_\Omega$, discrete Fourier transform (DFT) $\left(\mathcal{F}_Nc_{\underline{a}}\right)_A\in V_A$ is defined as
\footnote{We consider that $\left(\mathcal{F}_Nc_{\underline{a}}\right)_A$ means ``the index of $\mathcal{F}_Nc$ is $\underline{a}\in A$'', i.e., $\mathcal{F}_Nc_{\underline{a}}=\left(\mathcal{F}_Nc\right)_{\underline{a}}$ and not $\mathcal{F}_N\left(c_{\underline{a}}\right)$.\label{index a}}
\begin{equation}\label{DFT}
\left(\mathcal{F}_Nc_{\underline{a}}:=\sum_{\underline{\omega}\in\Omega}c_{\underline{\omega}}\underline{\omega}^{\underline{a}}\right)_A\in V_A,
\end{equation}
where $\underline{\omega}^{\underline{a}}$ is defined by $\omega_1^{a_1}\cdots\omega_N^{a_N}$, and $\omega^a$ is considered as the substituted value $\omega^a:=\left.x^a\right|_{x=\omega}$, that is, $\omega^a:=$1 for all $\omega\in\mathbb{F}_q$ if $a=0$.

\begin{Example}\rm
The simplest case is $N:=1$.
Note that, if $a\not=0$ and $\omega=0$, then $\omega^a=0$ trivially holds.
Thus $\left(\mathcal{F}_1c_a\right)_A\in V_A$ can be directly written as
$$
\mathcal{F}_1c_{a}=
\left\{
{\renewcommand\arraystretch{1.5}
\begin{array}{ll}
\sum_{\omega\in\Omega}c_\omega\omega^a=\sum_{\omega\in\Omega,\,\omega\not=0}c_\omega\omega^a & a\not=0\\
\sum_{\omega\in\Omega}c_\omega & a=0.
\end{array}}\right.
$$
Assume the next simplest case $N:=2$.
Then, for each $\left(a_1,a_2\right)=\left(a,b\right)\in A$, $\mathcal{F}_2c_{ab}$ can be directly written as
\begin{equation*}
\mathcal{F}_2c_{ab}=
\left\{
{\renewcommand\arraystretch{1.4}
\begin{array}{ll}
\sum_{(\psi,\omega)\in\Omega}c_{\psi\omega}\psi^{a}\omega^{b} & ab\not=0\\
\sum_{(\psi,\omega)\in\Omega,\,\psi\not=0}c_{\psi\omega}\psi^{a} & a\not=0,\,b=0\\
\sum_{(\psi,\omega)\in\Omega,\,\omega\not=0}c_{\psi\omega}\omega^{b} & a=0,\,b\not=0\\
\sum_{(\psi,\omega)\in\Omega}c_{\psi\omega}
& a=b=0.
\end{array}}\right.
\end{equation*}
Assume $N:=3$.
Then, for each $\left(a_1,a_2,a_3\right)=\left(a,b,c\right)\in A$, $\mathcal{F}_3c_{abc}$ can be directly written as
\begin{equation*}
\mathcal{F}_3c_{abc}=
\left\{
{\renewcommand\arraystretch{1.3}
\begin{array}{ll}
\sum\limits_{(\phi,\psi,\omega)\in\Omega}c_{\phi\psi\omega}\phi^{a}\psi^{b}\omega^{c} & abc\not=0\\
\sum_{(\phi,\psi,\omega)\in\Omega}c_{\phi\psi\omega}\phi^{a}\psi^{b}
 & ab\not=0,\,c=0\\
\sum_{(\phi,\psi,\omega)\in\Omega}c_{\phi\psi\omega}\phi^{a}\omega^{c}
 & ac\not=0,\,b=0\\
\sum_{(\phi,\psi,\omega)\in\Omega}c_{\phi\psi\omega}\psi^{b}\omega^{c}
 & bc\not=0,\,a=0\\
\sum_{(\phi,\psi,\omega)\in\Omega}c_{\phi\psi\omega}\phi^{a}
 & a\not=0,\,b=c=0\\
\sum_{(\phi,\psi,\omega)\in\Omega}c_{\phi\psi\omega}\psi^{b}
 & b\not=0,\,a=c=0\\
\sum_{(\phi,\psi,\omega)\in\Omega}c_{\phi\psi\omega}\omega^{c}
 & c\not=0,\,a=b=0\\
\sum_{(\phi,\psi,\omega)\in\Omega}c_{\phi\psi\omega} & a=b=c=0.
\end{array}}\right.
\end{equation*}
In general, to write $\mathcal{F}_Nc_{\underline{a}}$ directly, $2^N$ equalities are required.
\hfill$\Box$
\end{Example}

On the other hand, for $\left(h_{\underline{a}}\right)_A\in V_A$, inverse discrete Fourier transform (IDFT) $\left(\mathcal{F}_N^{-1}h_{\underline{\omega}}\right)_\Omega\in V_\Omega$ is defined as follows.
\footnote{Similarly to footnote \ref{index a}, $\left(\mathcal{F}_N^{-1}h_{\underline{\omega}}\right)_\Omega$ means that $\mathcal{F}_N^{-1}h_{\underline{\omega}}=\left(\mathcal{F}_N^{-1}h\right)_{\underline{\omega}}$ and not that $\mathcal{F}_N^{-1}h_{\underline{\omega}}=\mathcal{F}_N^{-1}(h_{\underline{\omega}})$.\label{index omega}}
For each $\underline{\omega}\in\Omega$, a subset $I=\left\{i_1\,\cdots,i_m\right\}$ of $\left[1,N\right]$ is determined such that $\omega_{i_1}\cdots\omega_{i_m}\not=0$ and $\omega_{i}=0$ for all $i\in\left[1,N\right]\backslash I$.
Then, define
\begin{equation}\label{IDFT}
\begin{split}
\mathcal{F}_N^{-1}h_{\underline{\omega}}&:=
(-1)^m\sum_{l_1,\cdots,l_m=1}^{q-1}
\\
&\hspace{-7mm}\left\{
\sum_{J\subseteq[1,N]\backslash I_{\underline{\omega}}}(-1)^{|J|}h_{\underline{i}(I_{\underline{\omega}},J)}
\right\}
\omega_{i_1}^{-l_1}\cdots\omega_{i_m}^{-l_m},
\end{split}
\end{equation}
where $\underline{i}(I,J):=\left(b_1,\cdots,b_N\right)\in A$ and, for $1\le i\le N$,
\begin{equation}\label{index}
b_i:=\left\{
\begin{array}{ll}
l_i & i\in I\\
q-1 & i\in J\\
0 & i\in[1,N]\left\backslash\left(I\cup J\right)\right..
\end{array}\right.
\end{equation}
For example, if $\omega_{i_1}\cdots\omega_{i_N}\not=0$ for $\underline{\omega}=\left(\omega_1,\cdots,\omega_N\right)\in\Omega$, then $I$ is equal to $[1,N]$, there is only one choice of $J=\emptyset$, and in this case the definition \eqref{IDFT} implies 
$$
\mathcal{F}_N^{-1}h_{\underline{\omega}}=
(-1)^N\sum_{l_1,\cdots,l_N=1}^{q-1}
h_{(l_1,\cdots,l_N)}
\omega_{i_1}^{-l_1}\cdots\omega_{i_m}^{-l_m},
$$
in other words, $\mathcal{F}_N^{-1}$ agrees with $N$-dimensional inverse discrete Fourier transform if $\Omega$ is restricted to $\left(\mathbb{F}_q^\times\right)^N$.
In general, for each $\underline{\omega}\in\Omega$, $\mathcal{F}_N^{-1}h_{\underline{\omega}}$ is equal to a linear combination of inverse discrete Fourier transforms whose dimensions do not exceed $N$.

\begin{Example}\rm
Assume $N:=1$.
If $\omega\not=0\in\Omega$, then $I=\{1\}\subseteq[1,1]$, $J=\emptyset\subseteq[1,1]\backslash I=\emptyset$, and $\underline{i}(I,J)=l_1=:i$.
If $\omega=0\in\Omega$, then $I=\emptyset\subseteq[1,1]$, $J=\emptyset,\{1\}\subseteq[1,1]\backslash I=\{1\}$, and $\underline{i}(I,J)=0,q-1$, respectively.
Thus $\left(\mathcal{F}_1^{-1}h_\omega\right)_\Omega\in V_\Omega$ can be directly written as
$$
\mathcal{F}_1^{-1}h_{\omega}:=
\left\{
{\renewcommand\arraystretch{1.5}
\begin{array}{ll}
-\sum_{i=1}^{q-1}h_{i}\omega^{-i} & \omega\not=0\\
h_{0}-h_{q-1} & \omega=0.
\end{array}}\right.
$$
Assume $N:=2$.
For $\left(\omega_1,\omega_2\right)=(\psi,\omega)\in\Omega$, for example, if $\psi\omega\not=0$, then $I=\{1,2\}\subseteq[1,2]$, $J=\emptyset\subseteq[1,2]\backslash I=\emptyset$, and $\underline{i}(I,J)=(l_1,l_2)=:(i,j)$; if $\psi\not=0$ and $\omega=0$, then $I=\{1\}\subseteq[1,2]$, $J=\emptyset,\{2\}\subseteq[1,2]\backslash I=\{2\}$, and $\underline{i}(I,J)=\underline{i}(\{1\},J)=(i,0),(i,q-1)$, repsectively.
Thus $\left(\mathcal{F}_2^{-1}h_{\psi\omega}\right)_\Omega\in V_\Omega$ can be directly written as \eqref{F-12}.
\begin{figure*}
\begin{gather}\label{F-12}
\mathcal{F}_2^{-1}h_{\psi\omega}:=
\left\{
{\renewcommand\arraystretch{1.4}
\begin{array}{ll}
\sum_{i,j=1}^{q-1}h_{ij}\psi^{-i}\omega^{-j} & \psi\omega\not=0\\
-\sum_{i=1}^{q-1}(h_{i,0}-h_{i,q-1})\psi^{-i} & \psi\not=0,\,\omega=0\\
-\sum_{j=1}^{q-1}(h_{0,j}-h_{q-1,j})\omega^{-j} & \psi=0,\,\omega\not=0\\
h_{0,0}-h_{0,q-1}-h_{q-1,0}+h_{q-1,q-1}
& \psi=\omega=0.
\end{array}}\right.
\\
\label{F-13}
\mathcal{F}_3^{-1}h_{\phi\psi\omega}:=
\left\{
{\renewcommand\arraystretch{1.3}
\begin{array}{ll}
-\sum_{i,j,l=1}^{q-1}h_{ijl}\phi^{-i}\psi^{-j}\omega^{-l} & \phi\psi\omega\not=0\\
\sum_{i,j=1}^{q-1}(h_{i,j,0}-h_{i,j,q-1})\phi^{-i}\psi^{-j} & \phi\psi\not=0,\,\omega=0\\
\sum_{i,l=1}^{q-1}(h_{i,0,l}-h_{i,q-1,l})\phi^{-i}\omega^{-l} & \phi\omega\not=0,\,\psi=0\\
\sum_{j,l=1}^{q-1}(h_{0,j,l}-h_{q-1,j,l})\psi^{-j}\omega^{-l} & \psi\omega\not=0,\,\phi=0\\
-\sum_{i=1}^{q-1}(h_{i,0,0}-h_{i,q-1,0}-h_{i,0,q-1}+h_{i,q-1,q-1})\phi^{-i} & \phi\not=0,\,\psi=\omega=0\\
-\sum_{j=1}^{q-1}(h_{0,j,0}-h_{q-1,j,0}-h_{0,j,q-1}+h_{q-1,j,q-1})\psi^{-j} & \psi\not=0,\,\phi=\omega=0\\
-\sum_{l=1}^{q-1}(h_{0,0,l}-h_{q-1,0,l}-h_{0,q-1,l}+h_{q-1,q-1,l})\omega^{-l} & \omega\not=0,\,\phi=\psi=0\\
\begin{array}{l}
h_{0,0,0}-h_{0,0,q-1}-h_{0,q-1,0}-h_{q-1,0,0}\\
\;\quad+h_{0,q-1,q-1}+h_{q-1,0,q-1}+h_{q-1,q-1,0}-h_{q-1,q-1,q-1}
\end{array}
& \phi=\psi=\omega=0.
\end{array}}\right.
\end{gather}
\end{figure*}
Assume $N:=3$.
For $\left(\omega_1,\omega_2,\omega_3\right)=(\phi,\psi,\omega)\in\Omega$, for example, if $\phi\not=0,\,\psi=\omega=0$, then $I=\{1\}\subseteq[1,3]=\{1,2,3\}$.
Then $J\subseteq[1,3]\backslash I=\{2,3\}$ has four choices, i.e., $J=\emptyset,\{2\},\{3\},\{2,3\}$, and respectively, $\underline{i}(I,J)=\underline{i}(\{1\},J)=(i,0,0),(i,q-1,0),(i,0,q-1),(i,q-1,q-1)$.
Thus $\left(\mathcal{F}_3^{-1}h_{\phi\psi\omega}\right)_\Omega\in V_\Omega$ can be directly written as \eqref{F-13}.
In general, the summand in each condition of $\underline{\omega}$ consists of $2^{N-m}$ terms, where $m$ is the number of nonzero components in $\underline{\omega}$.
\hfill$\Box$
\end{Example}

\begin{Proposition}\label{Fourier}
The two linear maps
\begin{align*}
&\left[V_A\ni\left(h_{\underline{a}}\right)_A\longmapsto\left(\mathcal{F}_N^{-1}h_{\underline{\omega}}\right)_\Omega\in V_\Omega\right]\\
&\left[V_\Omega\ni\left(c_{\underline{\omega}}\right)_\Omega\longmapsto\left(\mathcal{F}_Nc_{\underline{a}}\right)_A\in V_A\right]
\end{align*}
are inverse each other, that is, $\mathcal{F}_N\mathcal{F}_N^{-1}h_{\underline{a}}=h_{\underline{a}}$ and $\mathcal{F}_N^{-1}\mathcal{F}_Nc_{\underline{\omega}}=c_{\underline{\omega}}$.
\hfill$\Box$
\end{Proposition}

\subsection{Two vector spaces $V_S$ and $V_\Psi$
\label{vector spaces}}

Let $\Psi\subseteq\Omega$ and $n:=|\Psi|$.
\footnote{For any finite set $S$, the number of elements in $S$ is represented by $|S|$.}
One of the two vector spaces in the lemma is given by
$$
V_\Psi:=\left\{\left.\left(c_{\underline{\psi}}\right)_\Psi\,\right|\,\underline{\psi}\in\Psi,\,c_{\underline{\psi}}\in\mathbb{F}_q\right\},
$$
namely, $V_\Psi$ is the vector space over $\mathbb{F}_q$ indexed by the elements of $\Psi$, whose dimension is trivially $n$.
The other of the two vector spaces is somewhat complicated to define, since it requires Gr\"obner basis theory.
Let $\mathbb{F}_q[\underline{x}]$ be the ring of polynomials with coefficients in $\mathbb{F}_q$ whose variables are $x_1,\cdots,x_N$.
Let $Z_\Psi$ be an ideal of $\mathbb{F}_q[\underline{x}]$ defined by
$$
Z_\Psi:=\left\{\left.f(\underline{x})\in\mathbb{F}_q[\underline{x}]\,\right|\,f(\underline{\psi})=0\mbox{ for all }\underline{\psi}\in\Psi\right\}.
$$
We fix a monomial order $\preceq$ of $\left\{\left.\underline{x}^{\underline{s}}\,\right|\,\underline{s}\in\mathbb{N}_0^N\right\}$ \cite{BMS}.
We denote, for $f\not=0\in\mathbb{F}_q[\underline{x}]$,
\begin{gather*}
\mathrm{LM}(f):=\max_{\preceq}\left\{\left.\underline{x}^{\underline{s}}\,\right|\,\underline{s}\in\mathbb{N}_0^N,\,f_{\underline{s}}\not=0\right\}
\\
\mbox{if}\quad f=\sum_{\underline{s}\in\mathbb{N}_0^N,\,f_{\underline{s}}\not=0}f_{\underline{s}}\underline{x}^{\underline{s}}\in\mathbb{F}_q[\underline{x}],
\end{gather*}
whrere $\underline{x}^{\underline{s}}:=x_1^{s_1}\cdots x_N^{s_N}$ for $\underline{s}=\left(s_1,\cdots,s_N\right)\in\mathbb{N}_0^N$, and $\mathrm{LM}(f)$ is called the leading monomial of $f(\underline{x})\in\mathbb{F}_q[\underline{x}]$.
Then the support $S_\Psi=S\subseteq\mathbb{N}_0^N$ of $Z_\Psi$ for $\Psi$ is defined by
$$
S_\Psi=S:=\left.\mathbb{N}_0^N\right\backslash\left\{\mathrm{mdeg}\left(\mathrm{LM}(f)\right)\,\left|\,f(\underline{x})\in Z_\Psi\right.\right\},
$$
where $\mathrm{mdeg}\left(\underline{x}^{\underline{s}}\right):=\underline{s}\in\mathbb{N}_0^N$.
Fortunately, $S_\Psi$ has an intuitive description if a Gr\"obner basis $\mathcal{G}_\Psi$ of $Z_\Psi$ is obtained; it corresponds to the area surrounded by $\mathrm{LM}\left(\mathcal{G}_\Psi\right)$.
The support $S_\Psi=S\subseteq\mathbb{N}_0^N$ of $Z_\Psi$ for $\Psi$ is equivalently defined by
\begin{equation}\label{delta set}
\begin{split}
&\left.\left\{\underline{x}^{\underline{s}}\,\right|\,
\underline{s}\in S_\Psi\right\}
=\\
&\quad\left.\left\{\underline{x}^{\underline{s}}\,\left|\,
\underline{s}\in\mathbb{N}_0^N
\right.\right\}
\right\backslash
\left\{\mathrm{LM}(f)\,\left|\,
f(\underline{x})\in Z_\Psi
\right.\right\}.
\end{split}
\end{equation}
Then the other of the two vector spaces is given by
$$
V_S=V_{S_\Psi}:=\left\{\left.\left(h_{\underline{s}}\right)_S=\left(h_{\underline{s}}\right)_{S_\Psi}\,\right|\,\underline{s}\in S_\Psi,\,h_{\underline{s}}\in\mathbb{F}_q\right\},
$$
namely, $V_S$ is the vector space over $\mathbb{F}_q$ indexed by the elements of $S_\Psi$.
Since $\left.\left\{\underline{x}^{\underline{s}}\,\right|\,\underline{s}\in S_\Psi\right\}$ is a basis of $\mathbb{F}_{q}[\underline{x}]/Z_\Psi$ that is the quotient ring viewed as a vector space over $\mathbb{F}_{q}$, $V_S$ is isomorphic to $\mathbb{F}_{q}[\underline{x}]/Z_\Psi$.
It is known \cite{{Fitzgerald-Lax}},\cite{footprint} that the evaluation map
\begin{equation}\label{evaluation}
\mathbb{F}_{q}[\underline{x}]/Z_\Psi\ni
f\left(\underline{x}\right)\longmapsto
\left(f\left(\underline{\psi}\right)\right)_\Psi
\in V_\Psi
\end{equation}
is an isomorphism between two vector spaces.
Thus the map \eqref{evaluation} is also written as
\begin{equation}\label{ev}
V_S\ni
\left(h_{\underline{s}}\right)_S\longmapsto
\left(\sum_{\underline{s}\in S}h_{\underline{s}}\underline{\psi}^{\underline{s}}\right)_\Psi
\in V_\Psi,
\end{equation}
which is denoted as $\mathrm{ev}:V_S\to V_\Psi$.
In particular, it follows from the isomorphism \eqref{evaluation} or \eqref{ev} that $\left|S_\Psi\right|=|\Psi|$ and $\dim_{\mathbb{F}_q}V_S=n$.

Since $V_S$ and $V_\Psi$ have the same dimension $n$, it is trivial that $V_S$ is isomorphic to $V_\Psi$ as a vector space over $\mathbb{F}_q$.
However, this type of isomorphic maps depends on the choices of the bases of vector spaces; in addition, the normal orthogonal basis is not always convienient for encoding and decoding.
Our lemma asserts that there is a canonical isomorphism that does not depend on the bases.
As explained in Introduction, the isomorphic map $V_S\to V_\Psi$ of the lemma is the composition map of the extension and IDFT, which are defined accurately in the next subsection \ref{Isomorphic map}.
On the other hand, the inverse map  $V_\Psi\to V_S$ can be written concisely; that is ``DFT''
\begin{equation}\label{partial DFT}
V_\Psi\ni
\left(c_{\underline{\psi}}\right)_\Psi\longmapsto
\left(\sum_{\underline{\psi}\in\Psi}c_{\underline{\psi}}\underline{\psi}^{\underline{s}}\right)_S
\in V_S,
\end{equation}
which is actually the compound of DFTs in various dimensions.
It is shown from the definitions that the matrices that represent two maps \eqref{ev} and \eqref{partial DFT} are transposed each other if the bases of vector spaces are fixed.

\subsection{Isomorphic map $V_S\Carrow V_\Psi$
\label{Isomorphic map}}

Let $\mathcal{G}_\Psi$ be a Gr\"obner basis for the ideal $Z_\Psi$ with respect to $\preceq$.
We assume that $\mathcal{G}_\Psi$ consists of $d+1$ elements $\{g^{(u)}\}_{0\le u\le d}$, where
\begin{equation}\label{grobner}
\begin{split}
&g^{(u)}=g^{(u)}(\underline{x})=\\
&\underline{x}^{\underline{s}_u}+\sum_{\underline{s}\in S_\Psi}g_{\underline{s}}^{(u)}\underline{x}^{\underline{s}}\in\mathbb{F}_q[\underline{x}]
\;\mbox{ with }\underline{s}_u\in\left.\mathbb{N}_0^N\right\backslash S_\Psi.
\end{split}
\end{equation}
\begin{figure*}
\begin{align}\label{code}
C(R,\Psi)&:=\left\{\left(c_{\underline{\psi}}\right)_\Psi\in V_\Psi\left|\;
c_{\underline{\psi}}=\sum_{\underline{r}\in R}h_{\underline{r}}\underline{\psi}^{\underline{r}}\,
\mbox{ for some }\left(h_{\underline{r}}\right)_R\in V_R
\right.\right\}\\
C^\perp(R,\Psi)&:=\left\{\left(c_{\underline{\psi}}\right)_\Psi\in V_\Psi\left|\;
\sum_{\underline{\psi}\in\Psi}c_{\underline{\psi}}\underline{\psi}^{\underline{r}}=0
\mbox{ for all }\underline{r}\in R
\right.\right\}\label{dual code}
\end{align}
\end{figure*}
\begin{Definition}\label{recurrence}
We define that $\left(h_{\underline{a}}\right)_{A}\in V_A$ satisfies the linear recurrence relation from $\mathcal{G}_\Psi$ if and only if there exists $\left(h_{\underline{s}}\right)_{S}\in V_S$ such that, for all $\underline{a}\in A$ and all $0\le u\le d$,
\begin{equation}\label{relation}
h_{\underline{a}}+\sum_{\underline{s}\in S_\Psi}g_{\underline{s}}^{(u)}h_{\underline{a}+\underline{s}-\underline{s}_u}=0,
\end{equation}
where the indices $\underline{i}=\left(i_1,\cdots,i_N\right)$ of $h_{\underline{i}}$ are viewed within $1\le(i_l\,\mathrm{mod}\,(q-1))<q$ if $i_l\not=0$ for $1\le l\le N$.
Then we denote that $\left(h_{\underline{a}}\right)_A\hookleftarrow\left(h_{\underline{s}}\right)_S$.
\hfill$\Box$
\end{Definition}
Namely, each $h_{\underline{a}}$ satisfies $d+1$ equations.
Then we also say that $\left(h_{\underline{a}}\right)_{A}$ is the extension of $\left(h_{\underline{s}}\right)_S$.
In fact, there is one-to-one correspondence between arbitrary vectors $\left(h_{\underline{s}}\right)_S$ and all vectors $\left(h_{\underline{a}}\right)_{A}$ that satisfy the linear recurrence relation from $\mathcal{G}_\Psi$; from a given $\left(h_{\underline{s}}\right)_S$, generate $(h_{\underline{a}})_A$ inductively by
\begin{equation}\label{first generation}
h_{\underline{a}}:=
-\sum_{\underline{s}\in S_\Psi}
g_{\underline{s}}^{(u)}h_{\underline{a}+\underline{s}-\underline{s}_u}
\;\mbox{ for }\underline{i}\in\left.A\right\backslash S_\Psi.
\end{equation}
Then we obtain $\left(h_{\underline{a}}\right)_{A}$ that satisfies \eqref{relation}; the resulting values do not depend on the order of the generation because of the minimal property of Gr\"obner bases.
Conversely, from a given $\left(h_{\underline{a}}\right)_{A}$ that satisfies \eqref{relation}, we obtain a vector $\left(h_{\underline{s}}\right)_S$ by restricting $A$ to $S$.
Thus all $\left(h_{\underline{a}}\right)_{A}$ that satisfy the linear recurrence relation from $\mathcal{G}_\Psi$ are the extension of $\left(h_{\underline{s}}\right)_S$ by \eqref{first generation}.
Denote $\mathcal{E}:V_S\to V_A$ as the extension map $\left[V_S\ni\left(h_{\underline{s}}\right)_S\hookrightarrow\left(h_{\underline{a}}\right)_A\in V_A\right]$, and moreover, denote $\mathcal{R}:V_\Omega\to V_\Psi$ as the restriction map $\left[V_\Omega\ni\left(c_{\underline{\omega}}\right)_{\Omega}\mapsto\left(c_{\underline{\psi}}\right)_{\Psi}\in V_\Psi\right]$.
The following lemma is frequently used in this paper.
\begin{Main}
If $\underline{\omega}\in\left.\Omega\right\backslash\Psi$, it holds, for $\left(h_{\underline{a}}\right)_A\in\mathcal{E}\left(V_S\right)$, that $\mathcal{F}_N^{-1}h_{\underline{\omega}}=0$.
Moreover, the composition map $V_S\to V_\Psi$ in the following commutative diagram
\def\ext{\mathcal{E}}
\def\rest{\mathcal{R}}
\def\comp{\mathcal{C}}
\def\Fti{\mathcal{F}_N^{-1}}
\begin{diagram}
V_A & & \rTo^\Fti & & V_{\Omega} \\
\uTo^\ext & &  & & \dTo_\rest \\
V_S & & \rTo^\comp & & V_\Psi \\
\end{diagram}
gives an isomorphism between $V_S$ and $V_\Psi$.
The composition map is written as $\mathcal{C}:V_S\to V_\Psi$.
\hfill$\Box$
\end{Main}
Note that, if we admit that $V_A$ is isomorphic to $V_{\Omega}$ by $\mathcal{F}_N^{-1}$, then the first assertion of the lemma ``$\mathcal{F}_N^{-1}h_{\underline{\omega}}=0$ for $\underline{\omega}\not\in\Psi$'' deduces the isomorphism $V_S$ and $V_\Psi$ since the image of $\mathcal{E}\left(V_S\right)$ by $\mathcal{F}_N^{-1}$ agrees with $V_\Psi$.
We apply this lemma by putting $\Psi$ as the set of rational points, the set of erasure-and-error locations, and the set of redundant locations of codewords.

\section{Applications of main lemma
\label{Application}}

\subsection{Affine variety codes \cite{Fitzgerald-Lax}
\label{arbitrary subset}}

Let $\Psi\subseteq\Omega$ and $R\subseteq S_\Psi$.
Consider two types \eqref{code}, \eqref{dual code} of affine variety codes \cite{Fitzgerald-Lax} with code length $n:=|\Psi|$, where $\underline{\psi}^{\underline{r}}:=\psi_1^{r_1}\cdots\psi_N^{r_N}$ is defined samely as in \eqref{DFT}.
It follows from the isomorphic map $\mathrm{ev}:V_S\to V_\Psi$ of \eqref{ev} that
\begin{equation}\label{ev image}
C(R,\Psi)=\mathrm{ev}\left(V_R\right)
\end{equation}
and that $\left.\left\{\left(\underline{\psi}^{\underline{r}}\right)_\Psi\,\right|\,\underline{r}\in R\right\}$ is a linearly independent basis of $C(R,\Psi)$.
Since $\sum_{\underline{\psi}\in\Psi}c_{\underline{\psi}}\underline{\psi}^{\underline{r}}$ in \eqref{dual code} is the value of the inner product for $\left(c_{\underline{\psi}}\right)_\Psi$ and $\left(\underline{\psi}^{\underline{r}}\right)_\Psi$ in $V_\Psi$, the dual code of $C(R,\Psi)$ is equal to $C^\perp(R,\Psi)$.
Thus the dimension or the number of information symbols $k$ of $C^\perp(R,\Psi)$ is equal to $n-|R|$, in other words, $n-k=|R|$.

Consider a subspace $V_{S\backslash R}$ of $V_S$ with $S:=S_\Psi$ that has dimension $n-|R|$.
It follows from the isomorphic map $\mathcal{C}:V_S\to V_\Psi$ of the lemma that
\begin{equation}\label{C image}
C^\perp(R,\Psi)=\mathcal{C}\left(V_{S\backslash R}\right),
\end{equation}
which is similar to \eqref{ev image}.
While the definition \eqref{dual code} of $C^\perp(R,\Psi)$ is indirect and not constructive, the equality \eqref{C image} provides a direct construction and it corresponds to a non-systematic encoding of $C^\perp(R,\Psi)$.
Moreover, it is shown in the next subsection that the lemma also gives the systematic encoding for a class of such codes.

It is shown \cite{Fitzgerald-Lax} that $C(R,\Psi)$ and $C^\perp(R,\Psi)$ represent all linear codes over $\mathbb{F}_q$ respectively.
Futhermore, the decoding algorithm \cite{Fitzgerald-Lax} using Gr\"obner basis up to half the minimum distance $\lfloor(d_\mathrm{min}-1)/2\rfloor$ is shown for $C^\perp(R,\Psi)$.
However, since this type of decoding belongs to the class of NP-complete problems \cite{Berlekamp-McEliece-van Tilborg}, it is strongly suggested that the algorithm in \cite{Fitzgerald-Lax} does not run in polynomial time.

There is another algorithm \cite{Pellikaan} that decode all linear codes by $t$-error locating pair and solving system of linear equations.
The algorithm \cite{Pellikaan} can correct at least up to half the Feng--Rao minimum distance bound $\lfloor(d_\mathrm{FR}-1)/2\rfloor$ and its computational complexity equals O$\left(n^3\right)$, where $n$ is the code length of the linear code and $f(n)=\mathrm{O}\left(g(n)\right)$ means that $|f(n)|\le c|g(n)|$ for all $n$ and some constant $c>0$.

It is also shown \cite{Pellikaan-Shen-van Wee} that all linear codes over $\mathbb{F}_q$ are represented as algebraic geometric (AG) codes from algebraic curves.
As for fast decoding, Sakata {\it et al.}~\cite{Sakata-Jensen-Hoholdt} showed fast algorithm for decoding up to $\lfloor(d_\mathrm{FR}-1)/2\rfloor$ applicable to AG codes of one-point type, a well-studied subclass of AG codes.
Sakata {\it et al.}~\cite{Sakata-erasure} also showed that the similar algorithm to that in \cite{Sakata-Jensen-Hoholdt} can decode erasures and errors up to $\lfloor(d_\mathrm{FR}-1)/2\rfloor$ for one-point AG codes.
O'Sullivan \cite{generic},\cite{BMS} generalized BMS algorithm for finding the Gr\"obner basis of error locator ideal of affine variety codes.
However, fast decoding of affine variety codes including finding error values has been an open problem so far.

\subsection{DFT erasure-and-error decoding and systematic encoding
\label{DFT encoding}}

Consider the encoding problem for $C^\perp(R,\Psi)$.
From the lemma, non-systematic encoding is obtained as follows.
For $\left(h_{\underline{s}}\right)_S\in V_R$, let $\left(h_{\underline{a}}\right)_A\in\mathcal{E}\left(V_R\right)$ be its extended vector.
Then $\left(c_{\underline{\psi}}:=\mathcal{F}_N^{-1}h_{\underline{\psi}}\right)_\Psi\in C^\perp(R,\Psi)$ holds since
$\mathcal{F}_Nc_{\underline{s}}=\mathcal{F}_N\mathcal{F}_N^{-1}h_{\underline{s}}=h_{\underline{s}}=0$ for all $\underline{s}\in R$.
However, since error-correcting codes are usually encoded systematically,
it is natural to consider the systematic encoding for $C^\perp(R,\Psi)$, which is a certain type of erasure-only decoding as we observe in this subsection.

Let $\Phi\subseteq\Psi\subseteq\Omega$ so that $\Phi$ corresponds to the set of redundant positions and $\Psi\backslash\Phi$ corresponds to the set of information positions.
Then $S_\Phi\subseteq S_\Psi$ holds since $Z_\Phi\supseteq Z_\Psi$ and the definition \eqref{delta set}.
From now on, consider the linear codes $C:=C^\perp(S_\Phi,\Psi)$, i.e.,
\begin{equation}\label{the code}
\begin{split}
&C:=C^\perp(S_\Phi,\Psi)=\\
&\left\{\left(c_{\underline{\psi}}\right)_\Psi\in V_\Psi\,\left|\;
\sum_{\underline{\psi}\in\Psi}c_{\underline{\psi}}\underline{\psi}^{\underline{s}}=0
\mbox{ for all }\underline{s}\in S_\Phi
\right.\right\}.
\end{split}
\end{equation}
We choose $\Phi\subseteq\Psi$ such that $\Phi\not=\Psi$.
Then $k:=\dim_{\mathbb{F}_q}C=n-|\Phi|>0$ holds.

Suppose that erasure-and-error $\left(e_{\underline{\psi}}\right)_\Psi$ has occurred in a received word $\left(r_{\underline{\psi}}\right)_\Psi=\left(c_{\underline{\psi}}\right)_\Psi+\left(e_{\underline{\psi}}\right)_\Psi$ from the channel.
Let $\Phi_1\subseteq\Psi$ be the set of erasure locations and let $\Phi_2\subseteq\Psi$ be the set of error locations; we suppose that $\Phi_1$ is known but $\Phi_2$ and $\left(e_{\underline{\psi}}\right)_\Psi$ are unknown, that $e_{\underline{\psi}}\not=0\Rightarrow\underline{\psi}\in\Phi_1\cup\Phi_2$, and that $\underline{\psi}\in\Phi_2\Rightarrow e_{\underline{\psi}}\not=0$.
If $u+2t<d_\mathrm{FR}$ with $u:=|\Phi_1|$ and $t:=|\Phi_2|$ holds, where $d_\mathrm{FR}$ is the Feng--Rao minimum distance bound \cite{Andersen-Geil},\cite{Miura affine}, then it is known that the erasure-and-error version \cite{Koetter},\cite{Sakata-erasure} of Berlekamp--Massey--Sakata (BMS) algorithm \cite{generic},[\ref{BMS}, Chapter 10] or multidimensional Berlekamp--Massey (BM) algorithm calculates the Gr\"obner basis $\mathcal{G}_{\Phi_1\cup\Phi_2}$.
Since $\Phi_1$ is known, $\mathcal{G}_{\Phi_1}$ can be calculated by the ordinary error-only version in advance and then $\mathcal{G}_{\Phi_1\cup\Phi_2}$ can be calculated by the erasure-and-error version from the syndrome and the initial value $\mathcal{G}_{\Phi_1}$.
By using the recurrence from $\mathcal{G}_{\Phi_1\cup\Phi_2}$ and the lemma, the erasure-and-error decoding algorithm is realized as follows.

\begin{Algorithm}{\it DFT erasure-and-error decoding}
\label{DFT erasure-and-error}
@\setlength{\leftmargini}{4em}
\begin{description}
\setlength{\itemsep}{1mm}
\item[Input:]\ $\left(r_{\underline{\psi}}\right)_\Psi$ and $\Phi_1$
\item[Output:]\ \ $\left(c_{\underline{\psi}}\right)_\Psi\in C$
\item[Step 1.]\ $\left(h_{\underline{a}}\right)_A:=\left(\sum_{\underline{\psi}\in\Phi_1}\underline{\psi}^{\underline{a}}\right)_A$
\item[Step 2.]\ Calculate $\mathcal{G}_{\Phi_1}$ from syndrome $\left(h_{\underline{a}}\right)_A$
\item[Step 3.]\ $\left(\widetilde{r}_{\underline{s}}\right)_{S_\Phi}:=\left(\sum_{\underline{\psi}\in\Psi}r_{\underline{\psi}}\underline{\psi}^{\underline{s}}\right)_{S_\Phi}$
\item[Step 4.]\ Calculate $\mathcal{G}_{\Phi_1\cup\Phi_2}$ from $\left(\widetilde{r}_{\underline{s}}\right)_{S_\Phi}$ and $\mathcal{G}_{\Phi_1}$
\item[Step 5.]\ $\left(\widetilde{r}_{\underline{a}}\right)_A\hookleftarrow\left(\widetilde{r}_{\underline{s}}\right)_{S_\Phi}$ by $\mathcal{G}_{\Phi_1\cup\Phi_2}$
\item[Step 6.]\ $\left(e_{\underline{\psi}}\right)_\Psi:=\left(\mathcal{F}_N^{-1}\widetilde{r}_{\underline{\psi}}\right)_\Psi$
\item[Step 7.]\ $\left(c_{\underline{\psi}}\right)_\Psi:=\left(r_{\underline{\psi}}\right)_\Psi-\left(e_{\underline{\psi}}\right)_\Psi$
\hfill$\Box$
\end{description}
\end{Algorithm}

Next, we comment on the systematic encoding as erasure-only decoding.
Systematic means that, for a given information $\left(h_{\underline{\psi}}\right)_{\Psi\backslash\Phi}$, one finds $\left(c_{\underline{\psi}}\right)_\Psi\in C$ with $c_{\underline{\psi}}=h_{\underline{\psi}}$ for all $\underline{\psi}\in\Psi\backslash\Phi$.
Since $\Phi$ is known, systematic encoding can be viewed as an erasure-only decoding for $\left(e_{\underline{\psi}}\right)_\Phi:=\left(-c_{\underline{\psi}}\right)_\Phi$.
However, since $t:=0$ and $u:=n-k=|\Phi|$, the correctable erasure-and-error bound $u+2t<d_\mathrm{FR}$ does not hold in general.

Nevertheless we can show that systematic encoding works as an erasure-only decoding.
We calculate the Gr\"obner basis $\mathcal{G}_\Phi$ in advance, which has the role of generator polynomials in the case of RS codes.

\begin{Algorithm}{\it DFT systematic encoding}
\label{DFT systematic encoding}
@\setlength{\leftmargini}{4em}
\begin{description}
\setlength{\itemsep}{1mm}
\item[Input:]\ $\left(h_{\underline{\psi}}\right)_{\Psi\backslash\Phi}$ and $\Phi$
\item[Output:]\ \ $\left(c_{\underline{\psi}}\right)_\Psi\in C$ with $\left(c_{\underline{\psi}}\right)_{\Psi\backslash\Phi}=\left(h_{\underline{\psi}}\right)_{\Psi\backslash\Phi}$
\item[Step 1.]\ $\left(\widetilde{r}_{\underline{s}}\right)_{S_\Phi}:=\left(\sum_{\underline{\psi}\in\Psi\backslash\Phi}h_{\underline{\psi}}\underline{\psi}^{\underline{s}}\right)_{S_\Phi}$
\item[Step 2.]\ $\left(\widetilde{r}_{\underline{a}}\right)_{A}\hookleftarrow\left(\widetilde{r}_{\underline{s}}\right)_{S_\Phi}$ by $\mathcal{G}_{\Phi}$
\item[Step 3.]\ $\left(c_{\underline{\psi}}\right)_\Phi:=\left(-\mathcal{F}_N^{-1}\widetilde{r}_{\underline{\psi}}\right)_\Phi$
\hfill$\Box$
\end{description}
\end{Algorithm}

Thus the systematic encoding can be viewed as a special case of Algorithm \ref{DFT erasure-and-error} for $\left(r_{\underline{\psi}}\right)_\Psi=\left(h_{\underline{\psi}}\right)_\Psi$ with $h_{\underline{\psi}}:=0$ for $\underline{\psi}\in\Phi$.
Moreover, it can be obviously seen that any erasure-only $\left(e_{\underline{\psi}}\right)_{\Phi_1}$ can be decoded by Algorithm \ref{DFT erasure-and-error} if $S_{\Phi_1}\subseteq S_\Phi$.
It is expected that not only erasure-only but also erasure-and-error can be often decoded beyond the erasure-and-error correcting bound $u+2t<d_\mathrm{FR}$; in \cite{ISITA10}, the improvement and the necessary and sufficient condition that succeeds in erasure-and-error decoding are obtained for Hermitian codes.

\section{Estimation of complexity
\label{Estimation}}

We estimate the number of finite-field operations, i.e., additions, subtractions, multiplications, and divisions, in Algorithm \ref{DFT erasure-and-error} for codes \eqref{the code} since systematic encoding algorithm can be derived from Algorithm \ref{DFT erasure-and-error}.
Although there are various methods, e.g., fast Fourier transform (FFT), to reduce the manipulations in the algorithm, we mainly consider direct counting in their definitions because of conciseness.
Moreover, we enumerate not the strict numbers but approximate bounds.

Summarizing the results, we evaluate the algorithm as follows, where $n$ is code length, $N$ is dimension of $\Omega$, $q$ is finite-field size, $d$ is the number of elements in Gr\"obner bases.
$$
{\renewcommand\arraystretch{1.2}
\begin{tabular}{|c|c|}
\hline
Algorithm \ref{DFT erasure-and-error} & order of bound \\
\hline
Step 1 & $nNq^N$\\
\mbox{Step 2} & $dn^2$\\
\mbox{Step 3} & $n^2N$\\
\mbox{Step 4} & $dn^2$\\
\mbox{Step 5} & $nq^N$\\
\mbox{Step 6} & $nNq^N$\\
\mbox{Step 7} & $n$\\
\hline
\end{tabular}}
$$
Since $N\le d$, the total number of operations in Algorithm \ref{DFT erasure-and-error} has the order $dn^2+nNq^N$.
In the proof \cite{Fitzgerald-Lax} of $\{\mbox{linear codes}\}=\{\mbox{affine variety codes}\}$, $q^N$ is chosen as $q^N\ge n>q^{N-1}$, which leads $qn>q^N$ and $N\ge\log_q n>N-1$.
Thus $dn^2+nNq^N$ has the order of $n^2\left(d+q\log n\right)$.
Thus Algorithm \ref{DFT erasure-and-error} has the computational complexity of order $n^{2+\varepsilon}$, where $0\le\varepsilon<1$ and the implied constant depends on finite-field size $q$, which improves the order $n^3$ of that of Gaussian elimination for system of linear equations.

\section{Conclusion\label{Conclusion}}

In this paper, DFT and its inverse have been generalized to $\mathbb{F}_q^N$, and their Fourier inversion formula has been shown.
Moreover, a lemma for algebraic coding theory has been obtained.
As applications of our lemma, the construction of affine variety codes has been described, and fast erasure-and-error decoding and systematic encoding of a class of affine variety codes have been proposed.

\section*{Acknowledgment}
This work was supported in part by KAKENHI, Grant-in-Aid for Scientific Research (C) (23560478),
and was supported in part by a research grant from Storage Research Consortium (SRC).

\end{document}